\documentclass[10pt,twocolumn,letterpaper]{article}

\usepackage{btas}
\usepackage{times}
\usepackage{epsfig}
\usepackage{graphicx}
\usepackage{amsmath}
\usepackage{amssymb}
\usepackage{times}
\usepackage{epsfig}
\usepackage{graphicx}
\usepackage{amsmath}
\usepackage{amssymb}
\usepackage{verbatim}
\usepackage{booktabs} 
\usepackage{adjustbox}
\usepackage{colortbl} 
\usepackage{xcolor} 
\usepackage{xfrac}
\usepackage[]{algorithm2e}
\usepackage{subcaption}
\usepackage{graphicx,multirow}
\usepackage{array}
\btasfinalcopy % *** Uncomment this line for the final submission

\ifbtasfinal\fi

\begin{document}

%%%%%%%%% TITLE
\title{Prosodic-Enhanced Siamese Convolutional Neural Networks for Cross-Device Text-Independent Speaker Verification}

\author{Sobhan Soleymani, Ali Dabouei, Seyed Mehdi Iranmanesh, Hadi Kazemi, \\Jeremy Dawson, and Nasser M. Nasrabadi, {\it Fellow, IEEE}\\
West Virginia University\\
%Institution1 address\\
{\tt\small {\{ssoleyma, ad0046, seiranmanesh, hakazemi\}@mix.wvu.edu,}}\\
{\tt\small {\{jeremy.dawson, nasser.nasrabadi\}@mail.wvu.edu}}
% For a paper whose authors are all at the same institution,
% omit the following lines up until the closing ``}''.
% Additional authors and addresses can be added with ``\and'',
% just like the second author.
% To save space, use either the email address or home page, not both
%\and
%Second Author\\
%Institution2\\
%First line of institution2 address\\
%{\tt\small secondauthor@i2.org}
}

\maketitle
\thispagestyle{empty}
%%%%%%%%% ABSTRACT
\begin{abstract}
%In this paper a novel cross-device text-independent speaker verification architecture is proposed. Majority of the state-of-the-art deep architectures that are used for speaker verification tasks consider Mel-frequency cepstral coefficients. In contrast, our proposed Siamese convolutional neural network architecture uses Mel-frequency spectrogram coefficients to benefit from the dependency of the adjacent spectro-temporal features. Moreover, although spectro-temporal features have proved to be highly reliable in speaker verification models, the human voice consists of several linguistic levels such as acoustic, lexicon, prosody, and phonetics. The spectro-temporal features only represent some aspects of short-term acoustic level traits of the speaker's voice. To compensate for this shortcomings inherit in spectral features, we propose to enhance the Siamese convolutional neural network  architecture by deploying a multilayer perceptron network to incorporate the prosodic, jitter, and shimmer features. The proposed end-to-end verification architecture performs feature extraction and classification simultaneously. This proposed architecture displays significant improvement over classical signal processing approaches and deep algorithms for forensic cross-device speaker verification.
In this paper a novel cross-device text-independent speaker verification architecture is proposed. Majority of the state-of-the-art deep architectures that are used for speaker verification tasks consider Mel-frequency cepstral coefficients. In contrast, our proposed Siamese convolutional neural network architecture uses Mel-frequency spectrogram coefficients to benefit from the dependency of the adjacent spectro-temporal features. Moreover, although spectro-temporal features have proved to be highly reliable in speaker verification models, they only represent some aspects of short-term acoustic level traits of the speaker's voice. However, the human voice consists of several linguistic levels such as acoustic, lexicon, prosody, and phonetics, that can be utilized in speaker verification models. To compensate for these inherited shortcomings in spectro-temporal features, we propose to enhance the proposed Siamese convolutional neural network architecture by deploying a multilayer perceptron network to incorporate the prosodic, jitter, and shimmer features. The proposed end-to-end verification architecture performs feature extraction and verification simultaneously. This proposed architecture displays significant improvement over classical signal processing approaches and deep algorithms for forensic cross-device speaker verification. 
\end{abstract}
%%%%%%%%% BODY TEXT
\section{Introduction}
Speech is considered as a form of biometric verification, since everybody has his or her unique voice. Speaker verification aims to extract features from a speaker's speech samples and use them to recognize or verify speaker identity through modelings of the speaker's speech samples~\cite{liu2014speaker}. The speaker-verification literature focuses on designing a setup in which the claimed identity of a speaker is either accepted or rejected, which can be conducted as text-dependent~\cite{chen2015locally,variani2014deep,larcher2014text} or text-independent~\cite{heigold2016end,zhang2017end}. During text-dependent speaker verification the speech content is a predefined, fixed text, such as a passphrase, while text-independent speaker verification aims to verify the speaker using freeform spoken words, independent of the text or language or other prior constraints. The possible unconstrained variations in text-independent speaker verification make it much more challenging compared to text-dependent models~\cite{variani2014deep}.

Voice samples can be acquired through different recording devices and are subject to device and quality mismatch. In addition, the samples can be recorded at different sampling rates and distances, which result in bit-rate mismatch and channel noise. The samples are also subject to background noise problem due to environmental noise and distortion. Channel-independent speaker verification frameworks~\cite{nagrani2017voxceleb} try to address this problem. Channel-independent text-independent frameworks are considered to be the ultimate test in the speaker verification domain~\cite{nakasone2001forensic,heck2000robustness}.

Deep learning algorithms are the state-of-the-art frameworks for many biometric applications such as face~\cite{iranmanesh2018deep}, fingerprint~\cite{dabouei2018fingerprint}, and iris~\cite{Soleymani2018multi} classification, as well as multimodal classification~\cite{Soleymani2018multi,Soleymani2018generalized}, attribute-enhanced classification~\cite{kazemi2018attribute}, and domain adaptation~\cite{motiian2017few}.  
Deep learning architectures have recently proven to be able to provide superior performance compared to traditional speaker verification algorithms, showing significant gains over the state-of-the-art Gaussian Mixture Models and Hidden Markov Models~\cite{sainath2015deep,li2017deep,mclaren2015advances}. The majority of the deep learning architectures proposed for speaker recognition task are multilayer perceptron (MLP)-based models using Mel-frequency cepstral coefficients (MFCCs)~\cite{snyder2016deep,chen2015locally,variani2014deep}. However, MLP-MFCC architectures fail to preserve the correlation between the adjacent features. To address this issue convolutional neural networks (CNNs) are used in speaker recognition~\cite{zhang2017end,nagrani2017voxceleb}. Additionally, compared to architectures requiring hand-crafted features, convolutional neural networks (CNNs) extract and classify features simultaneously, and, therefore, avoid losing valuable information~\cite{krizhevsky2012imagenet}. 

State-of-the-art deep speaker recognition systems use spectro-temporal voice features~\cite{nagrani2017voxceleb,snyder2016deep}. The most well-known these short-term features used in the literature are spectrogram, MFCCs, and Mel-frequency spectrogram coefficients (MFSCs). Inheriting from the short-term nature of these features, most of the models proposed only explore the acoustic level of the signal, such as spectral magnitudes and formant frequencies~\cite{bhat2016recognition,farrus2007jitter}. However, several important linguistic levels such as lexicon, prosody or phonetics cannot be recognized from  short-term features. These levels of information are learned habits by the speaker. These features do not perform as well as the short-term features in the identification and verification scenarios when the utterances are significantly short. However, when the length of the utterance increases, it is shown that the identification and verification performance of the prosodic features increases drastically~\cite{park2017using}. These features also significantly improve the model when fused with short-term features~\cite{farrus2007jitter,farrus2006fusion}.

In the speaker-verification literature a three-phase procedure is defined. Initially, in the {\it development} phase, background models are developed from a large collection of data. New speakers are added to the model during the {\it enrollment} phase to construct speaker-dependent models. In the {\it evaluation} phase test utterances are compared to the enrolled speaker models and the background model to verify the identity of the speaker~\cite{variani2014deep}. In this setup, the difference between low-dimensional representations of enrollment and test utterances is considered to accept or reject the hypothesis~\cite{snyder2016deep}. However, in the proposed algorithm, the enrollment phase is excluded. The proposed Siamese model is trained using the utterances from the training set in the {\it training} phase. In the {\it test} phase, the trained model is deployed to compute the distance between two utterances. The computed inter sub-network distance is used to determine whether or not the utterances belong to the same speaker. 

In this paper, we make the following contributions: (i) prosodic, jitter, and shimmer features are deployed to enhance the performance of the proposed CNN Siamese network; (ii) a text-independent embedding space is constructed considering short-term and prosodic features; (iii) rather than extracting the features using hand-crafted methods, a fully data-driven architecture using fused CNN and MLP networks has been optimized for joint domain-specific feature extraction and representation with the application of speaker verification, finally (iv) the proposed algorithm can be used for real-time applications since it does not require the enrollment phase.

\section{Prosodic features to enhance deep coupled CNN}\label{sec:methodology}
CNN architectures have recently proven to outperform the traditional speaker verification algorithms. Following the scenario deployed in image processing literature, the input fed into the CNN is a nonlinearly scaled spectrogram with its first and second temporal derivatives~\cite{abdel2014convolutional}. CNN models prefer inputs that change smoothly along both dimensions. Therefore, acoustic features need to smoothly change both in time and frequency~\cite{abdel2012applying}. Since the acoustic signal is smooth in time, the frequency features need to preserve the locality of the speech signal. The majority of the works using deep neural networks for speech processing use MFCCs~\cite{snyder2016deep,variani2014deep,chen2011extracting}. 

However, these features do not preserve the locality of the frequency domain signal since the discrete cosine transform (DCT) projects the spectral energies into a basis that does not maintain locality~\cite{abdel2014convolutional}. Recently, MFSCs have been introduced to compensate for this shortcoming~\cite{abdel2012applying}. MFSCs are the log-energy computed directly from the mel-frequency spectral coefficients, which are the representation of the smoothed spectral envelope of the speech. These features, which are computed similarly to MFCC features with no DCT operation, along with their deltas and delta-deltas (first and second temporal derivatives) are fed into CNN as three channels of the input, describing the acoustic energy distribution of the spoken utterances.

These short-term coefficients represent the spectral envelope of a speech frame. Although these  parameters are speaker specific, they are unable to represent supra-segmental characteristics of the speech signal~\cite{farrus2007jitter}. On the other hand, prosodic coefficients represent features that are larger than phonetic units such as; sound, duration, tone and intensity variation. 

Although within-speaker variability in phonetic content and speaking style degrades the performance of speaker verification systems for short utterances~\cite{park2017using}, due to the practical complexity of the CNN architecture and the vast number of parameters that need to be trained, it is not feasible to feed the utterances to the network since it will drastically reduce the number of samples in the training set. To compensate for this shortcoming, we propose to compute the prosodic features from the whole utterances. For each utterance, several short utterances are randomly chosen. Each of these short utterances, along with the prosodic features calculated for the utterance, are fed to the network. The decision is made upon the computed overall scores. 

Following the setup in~\cite{farrus2007jitter}, 18 prosodic features are extracted from the utterances: three features related to word and segmental durations (number of frames per word and length of word-internal voiced and unvoiced segments), six features related to fundamental frequency (mean, maximum, minimum, range, pseudo-slope and slope), and nine jitter and shimmer measurements. Jitter indices used in this setup are absolute jitter, relative jitter, rap, and ppq5, while the shimmer indices used are shimmer (dB), relative shimmer, apq3, apq5, and apq11~\cite{farrus2007jitter}. 

Jitter and shimmer are defined as the indices for the cycle-to-cycle variations of fundamental frequency and amplitude, respectively. These indices are used to describe the voice quality. The frequency of a speaker's voice varies from one cycle to the next cycle. Jitter is defined as the cycle-to-cycle variation of fundamental frequency, and is the measurement of vocal stability. On the other hand, Shimmer is the index for vocal amplitude perturbation. Since these features characterize particular voices, they provide speaker-specific information.

\begin{figure*}
\begin{center}
\includegraphics[width=1\linewidth]{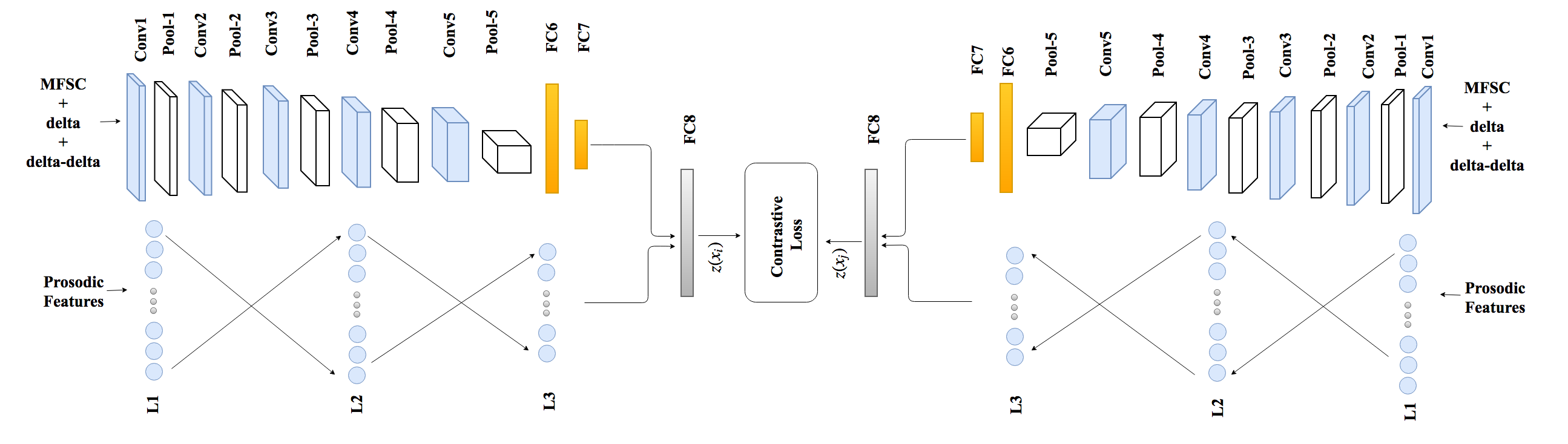}
\end{center}
\caption{Speaker verifier Siamese network. The MFSC-CNN consist of five convolutional and two fully-connected layers. The MLP network consists of two 64-units hidden layers and the output layer of $32$ units. This output layer along with FC7 layer are fed into FC8 of size $128$. Contrastive loss with weight-sharing is applied on sub-networks to compute the distance between two short utterances.}
\label{fig:1CNN}
\end{figure*}
\section{Proposed speaker-verification architecture}\label{sec:architecture}
The proposed Siamese architecture consists of two sub-networks that share weights. Each sub-network includes MLP and CNN networks, and the joint representation layer.  Segmental features are extracted from each utterance, and are fed to the MLP network, while random short utterances are chosen from the utterances. MFSCs are extracted from short utterances and fed into CNN network. Each sub-network is represented by a fully-connected fusion layer that act as the joint representation. Two joint representations are used to train the network through the contrastive loss. 
\subsection{Frequency- and prosody-domain networks}
Pooling algorithms are used in CNN architectures to reduce the possibility of over-fitting. {\it Maxpooling} is the sample-based conventional process in CNN architectures to down-sample the feature map representation without smoothing the feature maps, while extracting the most important features. It reduces the maps' dimensionality and allows the architecture to make sub-region assumptions. However, maxpooling is a shift-invariant operator and risks undesirable phonetic confusion~\cite{deng2013deep}. To compensate, we propose to use multiple maxpooling  sizes in the frequency domain instead of the conventional maxpooling and concatenate the output feature maps in depth as shown in Table~\ref{table:architecture_table}. %We have conducted d etailed error analysis on the effects of the CNN’s pool ing size. When a fixed pooling size increases f rom one to 12 , we found increasing confusions among the phones whose major formant freque ncies are close to each other; i n the meantime, the discrimination among phones whose formant frequencies are at extreme values tend s to impr ove. This observation and analysis motivated the development of a new type of pooling strategy in DNN s , which we describe below
%This is done to in part to help over-fitting by providing an abstracted form of the representation. As well, it reduces the computational cost by reducing the number of parameters to learn and provides basic translation invariance to the internal representation.
%Max pooling is done by applying a max filter to (usually) non-overlapping subregions of the initial representation.
On the other hand, since the proposed architecture is text-independent, the conventional last-layer fully-connected layer is replaced by average pooling along the time axis and a fully connected layer along the frequency axis. Additionally, this modification allows the inputs to vary in size in the time domain. 

The frequency domain network is comprised of five major convolutional components and two fully-connected layers which are connected in series. Each convolutional layer is followed by a rectified linear unit (ReLU) layer and a time domain maxpooling. $Conv2$ and $Conv4$ are also followed by a heterogeneous frequency domain maxpooling. In the proposed heterogeneous maxpooling, different kernel sizes are applied on the feature maps and the outputs are concatenated in depth and fed into the next convolutional layer.
%\subsection{Prosody-domain network}
The inputs to the frequency-domain network represent short-term features of the acoustic signal. 18 Prosodic features are fed into a multilayer perceptron with two hidden layers. Each  hidden layer consists of 64 hidden units, while the output layer includes 32 nodes. 
\begin{table}[t]
\caption[Table caption text]{The MFSC-dedicated CNN architectures. Notations (t) and (f) represent time and MFSC axis respectively.}  
\begin{center}
\addtolength{\tabcolsep}{-5pt}
\begin{tabular}{l@{\hskip .05in}c@{\hskip .05in}c@{\hskip .05in}c}
\toprule 
CNN-MFSC\\
\hline
layer&kernel&input&output\\
\hline
conv1& $3{\times} 3 {\times} 64$ &$40{\times} 300{\times} 3$& $40{\times} 300{\times} 64$\\
\rowcolor{black!10} maxpool1 (t)& $2 {\times} 1{\times}1$&$40{\times} 300{\times} 64$&$40{\times} 150{\times} 64$\\
conv2& $3{\times} 3 {\times} 128$&$40{\times} 150 {\times} 64$&$40{\times} 150 {\times} 128$\\
\rowcolor{black!10} maxpool2 (t)& $2 \times 1$&$40{\times} 150{\times} 128$&$40{\times} 75{\times} 128$\\
maxpool2-a (f)& $1 {\times}2{\times}1$&$40{\times} 75{\times} 128$&$20{\times} 75{\times} 128$\\
maxpool2-b (f)& $1 {\times} 3{\times}1$&$40{\times} 75{\times} 128$&$20{\times} 75{\times} 128$\\
maxpool2-c (f)& $1 {\times} 4{\times}1$&$40{\times} 75{\times} 128$&$20{\times} 75{\times} 128$\\
\rowcolor{black!10} conv3& $3{\times} 3 {\times} 256$ & $20{\times} 75 {\times} 384$& $20{\times} 75 {\times} 256$\\
 maxpool3 (t) & $2 {\times} 1{\times}1$ & $20 {\times} 75{\times}256$& $20 {\times} 37{\times}256$\\
\rowcolor{black!10} conv4 & $3{\times} 3 {\times} 256$ & $20{\times} 37 {\times} 256$& $20{\times} 37 {\times} 256$ \\
maxpool4 & $2 {\times} 1{\times} 1$& $20 \times 27 \times 256$& $20 \times 18\times 256$\\
\rowcolor{black!10} maxpool4-a (f)& $1 {\times} 2{\times}1$&$20{\times} 18{\times} 256$&$10{\times} 18{\times} 256$\\
\rowcolor{black!10}maxpool4-b (f)& $1 {\times} 3{\times}1$&$20{\times} 18{\times} 256$&$10{\times} 18{\times} 256$\\
\rowcolor{black!10}maxpool4-c (f)& $1 {\times} 4{\times}1$&$20{\times} 18{\times} 256$&$10{\times} 18{\times} 256$\\ 
conv5 & $3{\times} 3 {\times} 512$ & $10{\times} 18 {\times} 768$& $10{\times} 18 {\times} 512$\\
\rowcolor{black!10}avgpool (t) &  $1{\times} 18{\times} 1$ & $10{\times} 18{\times} 512$&$10{\times}1{\times} 512$\\
FC6 (f)&  $10{\times} 1{\times} 1024$ & $10{\times} 1{\times} 512$&$1{\times}1{\times} 1024$\\
\rowcolor{black!10}FC7&  $1{\times} 1\times 128$ & $1{\times} 1{\times} 1024$&$1{\times}1{\times} 128$\\
\bottomrule
\end{tabular}
\end{center}
\label{table:architecture_table}
\end{table}

%\begin{table}[h]
%\begin{center}
%\addtolength{\tabcolsep}{-5pt}
%\begin{tabular}{l@{\hskip .05in}c@{\hskip .05in}c@{\hskip .05in}c}
%\toprule 
%MLP-Prosodic\\
%\hline
%input(pool3)&$28\times 28\times 256$&$8\times 64\times 256$&$28\times 28\times 256$\\
%\toprule 
%layer&kernel&kernel&kernel\\
%\hline
%pool6& $4\times 4 $ & $4\times 4$& $4\times4$\\
%\rowcolor{black!10}FC7 &  $7\times 7\times 1024$ & $2\times 16\times 1024$&$7\times7\times 1024$\\
%\bottomrule
%\end{tabular}
%\end{center}
%\caption[Table caption text]{Added layers to each modality-dedicated network for multi-level feature abstraction fusion.}   
%\label{table:architecture_FC7}
%\end{table}
%\begin{figure*}[t]
%\begin{center}
%\includegraphics[width=1\linewidth]{figures/imageedit.png}
%\end{center}
%\caption{Joint representation architectures for BioCop(top) and BIOMDATA(bottom) databases.
%}
%\label{fig:fusion1m}
%\end{figure*}
%\subsection{Loss Function}
 
\subsection{Speaker-verification coupled CNN}
The final objective of the proposed model is to verify whether or not two utterances recorded on different devices belong to the same speaker or not. The utterances can also be recorded at the same time or in different sessions. Therefore, the proposed method needs to satisfy the text-independent  condition. On the other hand, it is not feasible to feed the whole utterances to the network, since it drastically reduces the number of samples. In addition, since the utterance can vary in length, feeding them to the network limits the batch normalization benefits. Therefore, we propose to randomly choose several fixed-length short utterances. Each short utterance is fed into the network along with the prosodic features calculated from the long utterance. The final decision is made upon the distances (scores) given to each pair of short utterances. 

As can be seen in Figure~\ref{fig:1CNN}, the proposed architecture is a Siamese network, where two sub-networks share weights. Each sub-network consists of CNN and MLP networks. The MFSC-CNN consist of five convolutional and two fully-connected layers. The MLP network consists of two 64-units hidden layers and the output layer of $32$ units. This output layer, along with the FC7 layer, are fed into a fully-connected layer of size $128$. Contrastive loss is applied to compute the distance between two short utterances. 

The ultimate goal of the proposed architecture is to find the latent deep features 
representing the speaker specific features. In order to find a common latent embedding subspace, we couple sub-networks via a contrastive loss function~\cite{chopra2005learning}. This function ($L_{c}$) pulls the utterances that belong to the same speaker toward each other into a common latent embedding subspace and pushes the utterances belong to different speakers apart.

Although the utterances came from different devices, the recording device is assumed unknown in the test process. Therefore, the sub-networks cannot be trained for a specific device. Considering no knowledge about the recording device, weight-sharing between sub-networks is assumed. The contrastive loss between the sub-networks is defined as~\cite{chopra2005learning}:
\begin{equation}
\small
L_{c}(x_i,x_j;y_{i,j})=(1-y_{i,j})l_{ge}(x_i,x_j)+y_{i,j}l_{im}(x_i,x_j),
\end{equation}
where $x_i$ and $x_j$ are two utterances. The binary label $y_{i,j}$ is equal to $0$ if $x_i$ and $x_j$ belong to the same speaker. Otherwise, it is equal to $1$.  $l_{ge}$ and $l_{im}$ represent the partial loss functions for the genuine and impostor pairs, respectively, and $D_{i,j}$ indicates the Euclidean distance between the embedded data in the common feature subspace (FC8). $l_{ge}$ and $l_{im}$ are defined as follows: 
\begin{equation}
\small
l_{ge}(x_i,x_j)=\frac{1}{2}  ||z(x_i)-z(x_j)||_2\quad \text{    for} \quad y_{i,j}=0,
\end{equation}
\begin{equation}
\small
l_{im}(x_i,x_j)=\frac{1}{2}max(0,m-||z(x_i)-z(x_j)||_2)\quad\text{for}\quad y_{i,j}=1,
\end{equation}
\noindent where $m$ is the contrastive loss margin. $z(x) $ is the sub-network based embedding functions, which transforms $x$ into the common latent embedding space. It should be noted that the contrastive loss function considers the subjects' labels inherently. Therefore, it has the ability to find a discriminative embedding space by employing the data labels in contrast to some other metrics, such as Euclidean distance. This discriminative embedding space would be useful in identifying speaker specific features. During the training phase, the pairs of the short-utterances are fed into the Siamese network along with the prosodic features computed from the pair of whole utterances. During the test phase, for a pair of utterances, first the prosodic features are computed. Then, several pairs of short utterances are randomly chosen and fed to the network. The distance for each pair is computed. The distance between two long utterances is defined as the mean of these distances.
\section{Joint optimization of the network}
In this section, the training of the Siamese architecture is discussed. Here, we explain the implementation of CNN and MLP networks, the joint fully-connected fusion layer and the concurrent optimization of the architecture. 
\subsection{Training of the network}
Initially, the MFSC-dedicated CNNs are trained independently as a classifier using all the utterances in the training set. As explained in Section~\ref{sec:architecture}, the network consists of five convolutional and two fully-connected layers. A softmax layer is added to the network, where the number of units is equal to the number of speakers in the training set. Training the network as classifier facilitates the extraction of the discriminative features from MFSC coefficients.

The inputs are $3$ seconds utterances which are represented as $300\times 224\times 3$ images. Three channels represent static, delta and delta-delta feature maps, while $40$ represent the number of MFSC coefficients. %  The preprocessing algorithm includes the channel-wise mean subtraction on RGB values, where channel means are calculated on the whole training set. 
The training algorithm is deployed by minimizing the softmax cross-entropy loss using mini-batch stochastic gradient descent with momentum. The training was regularized by weight decay and $50\%$ dropout for the fully connected layers except for the last layer. The batch size, momentum and L$_2$ penalty multiplier are set to 32, 0.9 and $0.0005$, respectively. The initial learning rate is set to $0.1$. The learning rate is decreased exponentially by a factor of $0.1$ for every $2$ epochs of training. In this network, batch normalization~\cite{ioffe2015batch} is applied. The moving average decay is set to $0.99$. 

Similarly, the MLP network is optimized independently. The parameters for this optimization are the same as the parameters for the CNN network. To train the joint representation, the CNN and MLP networks are frozen and the joint representation layer is optimized greedily upon the extracted features. %The optimization parameters are the same as the fingerprint network. Finally all networks are jointly optimized. Here, the batch size is further reduced %to $16$ 
The initial learning rate is reduced to the smallest final learning rate among two networks. Finally, the classification architecture is trained jointly.% and the corresponding joint representation network.    
%\subsection{Two-modality networks}
%We have also trained two-modality architectures. These networks can be used when one of the three modalities is missing. All three possible two-modality architectures are trained for both concatenated and bilinear feature fusion using the process explained in section~\ref{trainjointrepresentation}. 

To train the Siamese network, the network is initialized with the weights optimized for the classifier network. The pairs are fed into the network and the contrastive loss function is minimized while the sub-networks share weight.
\subsection{Hyperparameter optimization}
The hyperparameters in our experiments are : $\lambda$ the regularization parameter, $\alpha_0$ initial learning rate, $n$ number of epochs per decay for the learning rate, $d$ moving average decay, and $m$ as the momentum. For each optimization,  the 5-fold cross-validation method on the training set is used to estimate the best hyperparameters.
%\begin{figure*}[h]
%\begin{subfigure}{.5\textwidth}
%\begin{center}
%\includegraphics[width=1\linewidth]{figures/ICB2018_Rank_Biocop.png}
%\end{center}
%\caption{}
%\label{fig:fusion1}
%\end{subfigure}
%\begin{subfigure}{.5\textwidth}
%\begin{center}
%\includegraphics[width=1\linewidth]{figures/ICB2018_Rank_BIOMDATA.png}
%\end{center}
%\caption{}
%\label{fig:fusion2}
%\end{subfigure}
%\caption{ CMC curves for (a) BioCop, and  (b) BIOMDATA databases.}
%\label{fig:fusion2vs1}
%\end{figure*}
%\begin{figure*}[h]
%\begin{subfigure}{.5\textwidth}
%\begin{center}
%\includegraphics[width=1\linewidth]{figures/Rank_all_average1.png}
%\end{center}
%\caption{}
%\label{fig:fusion3}
%\end{subfigure}
%\begin{subfigure}{.5\textwidth}
%\begin{center}
%\includegraphics[width=1\linewidth]{figures/Rank_averages.png}
%\end{center}
%\caption{}
%\label{fig:fusion4}
%\end{subfigure}
%\caption{ The performance of three-modality networks compared with (a) single-modality networks and (b) the averages of the two-modality networks.}
%\label{fig:fusion3vs1}
%\end{figure*}
\section{Experiments and disscussions}\label{sec:Experiments}
%To compare the results for the proposed algorithms, with the state-of-the-art algorithms, Gabor features in five scales and eight orientations are extracted from all modalities. For the face images, $31,360$ features are extracted from $224\times 224$ aligned images. While, for the iris images, $36,630$ features are extracted from $64 \times 512$ segmented and normalized image. In the case of fingerprint images, $31,360$ features are extracted from the enhanced $224\times 224$ images, as described in Section~\ref{preprocessing}, around the core point.      
%\subsection{Evaluation metrics}\label{sec:evaluation}
%The performance of different experiments are reported and compared using two classification metrics.~The utilized metrics are classification accuracy and \emph{Recall@K}.~The accuracy is the fraction of correctly classified samples regarding their classes. The \emph{Recall@K} metric is the probability that a subject class is correctly classified at least at rank-k while the candidate classes are sorted by their similarity score to the query samples.~The calculation of \emph{Recall@K} is done per class and is averaged over all available classes. 
\subsection{Dataset}\label{Dataset}
%To evaluate the performance of the proposed weighted fusion multimodal architecture, two challenging multimodal biometric databases are considered.
{\bf {FBI Voice Collection 2016:}} This database consists of two sessions (July 2016 and
January 2017) of speech from 411 individuals using three recording devices: a high-
quality microphone, a typical interview room recording system/DVR, and a digital recorder capturing the speech over a cell phone connection. The last two recording are recorded simultaneously. The number of male and female speakers are $205$ and $206$ respectively. This database is one of the few databases that allows disjoint channel-independent training and testing of the proposed algorithm. The total number of utterances is equal to $2,418$. The training is conducted on 361 speakers. The test is performed on the remaining 50 subjects. 
  %that have samples in all six modalities. For each modality, four randomly chosen samples are used for training and the remaining samples are used for testing. For any modality that the number of the samples is less than five, one sample is used for testing and the remaining samples are used for training. 
 A summary of the database is presented in Table~\ref{table:datasetsFBI}. 
\begin{table}[h]
\caption[Table caption text]{The number of utterances in train and test sets for FBI voice collection 2016 database.}
\begin{center}
\begin{tabular}{ l| c| c}
& Train set &Test set \\
\hline
Utterances&2148&300\\
\hline
Microphone&705&100\\
DVR&708&100\\
Phone&705&100\\
\end{tabular}
\end{center}
\label{table:datasetsFBI}
\end{table}
%For both training and test sets, for each individual, $250$ set of samples are randomly chosen. Each set includes normalized left and right irises, and enhanced left index, right index, left thumb and right thumb fingerprint images. The total number of samples in the training and test sets are $54,750$.
\subsection{Data representation}\label{preprocessing}
Initially all utterances are re-sampled to $48$ KHz. For each utterance prosodic features were extracted using Praat software for acoustic analysis~\cite{pratt}. These 18 features listed in Section~2 are the inputs fed to MLP network. Then, voiced segments of the utterances are detected using the voicebox toolbox~\cite{VOICEBOX}. Each voiced utterance is divided into $25$ms frames with $60\%$ overlap. 

Each frame is multiplied with a hamming window to keep the continuity of the first and the last points in the frame. $40$ MFSC coefficients are extracted from each frame. Delta and delta-delta channels are constructed for each frame as the first derivative and second temporal derivative of MFSC features. Cepstral mean and variance normalization are applied on each utterance, in which each frequency bin is normalized to zero mean and unit variance. Finally, short utterances of three seconds length with two seconds overlap are generated. The inputs fed to CNN network are $40\times 300\times 3$ short utterances. 

\subsection{Training and test phases}\label{Training}
{\bf Training phase:} Pairs of short utterances are randomly chosen, while we make sure that the overall number of genuine and imposter pairs are equal. The pairs of short utterances are fed into the architecture along with the prosodic features. The architecture is trained under contrastive loss with no normalization on the last fully-connected layer (FC8). Here the short utterances are assumed to be independent samples, and the contrastive loss is applied on each pair of short utterances. The contrastive loss margin is set to $10$.

{\bf Test phase:} For each pair of utterances, $500$ pairs of short utterances are randomly chosen. The pairs of short utterances are fed into the architecture along with the prosodic features. For each pair of short utterances, the distance is computed as the Euclidean distance between the samples in the embedding space. The vector of the distances between the short utterances is used to determine the distance between two utterances. The short samples can be noisy or may not include speaker specific information. Therefore, averaging the distances between pairs of short utterances may include outliers. To remove the effect of these short utterances, the vector's mean and standard deviation are computed. The average of the elements in the vicinity of two standard deviations from the mean value represent the distance between the pair of utterances.
\vspace{-1mm}
\subsection{Evaluation metrics}\label{sec:evaluation}
The performance of different experiments are reported and compared using two verification metrics. The utilized metrics are equal error rate (EER) and area under curve (AUC). When false acceptance and false rejection rate for the model are equal, the common value is referred to as EER. AUC represents the area under the receiver operating characteristic curve. %The reported values are the average values for five randomly generated training and test sets. 
\vspace{-1mm}
\subsection{Results}\label{Results}
\vspace{-1mm}
Table~\ref{table:results} presents the verification results for the proposed algorithm. In addition, the verification results for CNN and MLP trained independently are presented. The score-level fusion of two networks is considered as well. The performance of the proposed algorithm is compared with that of i-vector/PLDA algorithm~\cite{dehak2011front}. The same MFSC feature used in the proposed deep algorithm are used in i-vector algorithm. The i-vector model is also trained with MFCC features. The algorithm is also compared with two state-of-the-art deep architectures~\cite{nagrani2017voxceleb,chen2015locally}.
\begin{table}[t]
\caption[Table caption text]{Verification performance on FBI Voice Collection 2016 database.}   
\begin{center}
%\addtolength{\tabcolsep}{-5pt}
\begin{tabular}{l|c|c}%{l@{\hskip .05in}c@{\hskip .05in}c@{\hskip .05in}c}
Algorithm & AUC&EER\\
\hline
i-vector/PLDA (MFSC)&0.9153&0.1579\\
i-vector/PLDA (MFCC)&0.9185&0.1526\\
Chen et al.~\cite{chen2015locally}&0.9207&0.1451\\
Nagrani et al.~\cite{nagrani2017voxceleb}&0.9215&0.1469\\
CNN network &0.9218&0.1421\\
Prosodic network &0.9011&0.1673\\
Score-level fusion&0.9148&0.1578\\
Coupled network &0.9358&0.1311\\
\end{tabular}
\end{center}
\label{table:results}
\end{table}
%\begin{table}[h]
%\begin{center}
%\begin{tabular}{c| l c c c}
%\toprule 
%\multicolumn{1}{c}{} & \multicolumn{1}{c}{} & \multicolumn{1}{c}{KNN} & \multicolumn{1}{c}{SVM} &\multicolumn{1}{c}{CNN}\\ \hline
%\hline
%\multirow{3}{*}{\rotatebox[origin=c]{90}{BioCop}}&Face&89.68&88.76&98.14\\
%&Iris&70.52&79.26&99.05\\
%&Right index&91.22&90.61&97.28\\
%\hline
%\multirow{6}{*}{\rotatebox[origin=c]{90}{BIOMDATA}}&Left iris &66.61&71.92&99.35\\
%&Right iris&64.89&71.08&98.95 \\
%&Left thumb&61.23&63.96 &80.15 \\
%&Left index&82.91&84.70 &93.43 \\
%&Right thumb&62.11&63.52&82.63\\
%&Right Index&82.05&84.46&93.12\\
%\bottomrule
%\end{tabular}
%\end{center}
%\caption[Table caption text]{Rank-one recognition rate for BioCop and BIOMDATA databases.}
%\label{table:results_single}
%\end{table}

%\subsection{Channel-dependent special case}\label{special_case}
Table~\ref{table:results_dependent} presents the results for channel-dependent setup. In this special case, each sub-network is fed with utterances from a specific device. To train this architecture, the sub-networks do not share weights. To initialize the parameters for this setup, both the sub-networks are initialized with the parameters from channel-independent setup. This setup leads to better performance compared to channel-independent setup, since the channel-dependent information in the test samples can be learned during the training phase. The only exception is Phone-DVR cross-device verification setup, where, both devices are considered low-quality devices.

\begin{table}[t]
\caption[Table caption text]{EER and AUC for the channel-dependent setup.}
\begin{subtable}[h]{0.5\textwidth}
\begin{center}
\addtolength{\tabcolsep}{-0pt}
\begin{tabular}{l|c|c|c}%{l@{\hskip .05in}c@{\hskip .05in}c@{\hskip .05in}c}
Device&Microphone&DVR&Phone\\
\toprule 
Microphone&0.0712&0.1132 &0.1247\\
\hline
DVR&\cellcolor{black!10}&0.0827 &0.2069\\
\hline
Phone&\cellcolor{black!10}&\cellcolor{black!10} &0.1316
\end{tabular}
\end{center}
\caption[Table caption text]{EER}   
\label{table:results_AUC}
\end{subtable}
%\end{table}
%\begin{table}[t]
%\small
\begin{subtable}[h]{0.5\textwidth}
\begin{center}
\addtolength{\tabcolsep}{-0pt}
\begin{tabular}{l|c|c|c}%{l@{\hskip .05in}c@{\hskip .05in}c@{\hskip .05in}c}
Device&Microphone&DVR&Phone\\
\toprule 
Microphone&0.9785&0.9537 &0.9512\\
\hline
DVR&\cellcolor{black!10}&0.9717 &0.8467\\
\hline
Phone&\cellcolor{black!10}&\cellcolor{black!10} &0.9547
\end{tabular}
\end{center}
\caption[Table caption text]{AUC}       
\label{table:results_BIOMDATA}
\end{subtable}
\label{table:results_dependent}
\end{table}
\section{Conclusion}
In this paper we proposed a novel cross-device text-independent speaker verification Siamese architecture, where Mel-frequency spectrogram coefficients are used to benefit from correlation of the adjacent features. In addition, prosodic features were deployed to enhance the spectral features fed to CNN. A MLP network is trained to represent the prosodic features describing words, fundamental frequency, jitter and shimmer. The joint representation fusing two networks, trains the network through contrastive loss. The proposed end-to-end verification architecture performs feature extraction and verification simultaneously. The proposed architecture displays significant improvement over conventional classical and deep algorithms for forensic cross-device speaker verification. 

\begin{center}
ACKNOWLEDGEMENT
\end{center}
This work is based upon a work supported by the Center for Identification Technology Research and the National Science Foundation under Grant $\#1650474$.

{\small
\bibliographystyle{ieee}
\bibliography{bib}

\begin{thebibliography}{10}\itemsep=-1pt

\bibitem{pratt}
Praat software: http://www.fon.hum.uva.nl/praat/.

\bibitem{VOICEBOX}
Voicebox: Speech processing toolbox for matlab:
  http://www.ee.ic.ac.uk/hp/staff/dmb/voicebox/voicebox.html.

\bibitem{abdel2014convolutional}
O.~Abdel-Hamid, A.-r. Mohamed, H.~Jiang, L.~Deng, G.~Penn, and D.~Yu.
\newblock Convolutional neural networks for speech recognition.
\newblock {\em IEEE/ACM Transactions on audio, speech, and language
  processing}, 22(10):1533--1545, 2014.

\bibitem{abdel2012applying}
O.~Abdel-Hamid, A.-r. Mohamed, H.~Jiang, and G.~Penn.
\newblock Applying convolutional neural networks concepts to hybrid nn-hmm
  model for speech recognition.
\newblock In {\em IEEE International Conference on Acoustics, Speech and Signal
  Processing (ICASSP)}, pages 4277--4280, 2012.

\bibitem{bhat2016recognition}
C.~Bhat, B.~Vachhani, and S.~K. Kopparapu.
\newblock Recognition of dysarthric speech using voice parameters for speaker
  adaptation and multi-taper spectral estimation.
\newblock In {\em Proc. Interspeechs}, pages 228--232, 2016.

\bibitem{chen2011extracting}
K.~Chen and A.~Salman.
\newblock Extracting speaker-specific information with a regularized siamese
  deep network.
\newblock In {\em Advances in Neural Information Processing Systems}, pages
  298--306, 2011.

\bibitem{chen2015locally}
Y.-h. Chen, I.~Lopez-Moreno, T.~N. Sainath, M.~Visontai, R.~Alvarez, and
  C.~Parada.
\newblock Locally-connected and convolutional neural networks for small
  footprint speaker recognition.
\newblock In {\em Annual Conference of the International Speech Communication
  Association}, 2015.

\bibitem{chopra2005learning}
S.~Chopra, R.~Hadsell, and Y.~LeCun.
\newblock Learning a similarity metric discriminatively, with application to
  face verification.
\newblock In {\em IEEE Conference on Computer Vision and Pattern Recognition
  (CVPR), 2005}, volume~1, pages 539--546, 2005.

\bibitem{dabouei2018fingerprint}
A.~Dabouei, H.~Kazemi, S.~M. Iranmanesh, J.~Dawson, and N.~M. Nasrabadi.
\newblock Fingerprint distortion rectification using deep convolutional neural
  networks.
\newblock In {\em International Conference on Biometrics}, 2018.

\bibitem{dehak2011front}
N.~Dehak, P.~J. Kenny, R.~Dehak, P.~Dumouchel, and P.~Ouellet.
\newblock Front-end factor analysis for speaker verification.
\newblock {\em IEEE Transactions on Audio, Speech, and Language Processing}.

\bibitem{deng2013deep}
L.~Deng, O.~Abdel-Hamid, and D.~Yu.
\newblock A deep convolutional neural network using heterogeneous pooling for
  trading acoustic invariance with phonetic confusion.
\newblock In {\em IEEE International Conference on Acoustics, Speech and Signal
  Processing (ICASSP)}, pages 6669--6673, 2013.

\bibitem{farrus2006fusion}
M.~Farr{\'u}s, A.~Garde, P.~Ejarque, J.~Luque, and J.~Hernando.
\newblock On the fusion of prosody, voice spectrum and face features for
  multimodal person verification.
\newblock In {\em Ninth International Conference on Spoken Language
  Processing}, 2006.

\bibitem{farrus2007jitter}
M.~Farr{\'u}s, J.~Hernando, and P.~Ejarque.
\newblock Jitter and shimmer measurements for speaker recognition.
\newblock In {\em Eighth Annual Conference of the International Speech
  Communication Association}, 2007.

\bibitem{heck2000robustness}
L.~P. Heck, Y.~Konig, M.~K. S{\"o}nmez, and M.~Weintraub.
\newblock Robustness to telephone handset distortion in speaker recognition by
  discriminative feature design.
\newblock {\em Speech Communication}, 31(2-3):181--192, 2000.

\bibitem{heigold2016end}
G.~Heigold, I.~Moreno, S.~Bengio, and N.~Shazeer.
\newblock End-to-end text-dependent speaker verification.
\newblock In {\em IEEE International Conference on Acoustics, Speech and Signal
  Processing (ICASSP)}, pages 5115--5119, 2016.

\bibitem{ioffe2015batch}
S.~Ioffe and C.~Szegedy.
\newblock Batch normalization: Accelerating deep network training by reducing
  internal covariate shift.
\newblock {\em arXiv preprint}, 2015.

\bibitem{iranmanesh2018deep}
S.~M. Iranmanesh, A.~Dabouei, H.~Kazemi, and N.~M. Nasrabadi.
\newblock Deep cross polarimetric thermal-to-visible face recognition.
\newblock {\em arXiv preprint arXiv:1801.01486}, 2018.

\bibitem{kazemi2018attribute}
H.~Kazemi, S.~Soleymani, A.~Dabouei, M.~Iranmanesh, and N.~M. Nasrabadi.
\newblock Attribute-centered loss for soft-biometrics guided face sketch-photo
  recognition.
\newblock In {\em IEEE Conference on Computer Vision and Pattern Recognition
  Workshop}, 2018.

\bibitem{krizhevsky2012imagenet}
A.~Krizhevsky, I.~Sutskever, and G.~E. Hinton.
\newblock Imagenet classification with deep convolutional neural networks.
\newblock In {\em Advances in neural information processing systems}, pages
  1097--1105, 2012.

\bibitem{larcher2014text}
A.~Larcher, K.~A. Lee, B.~Ma, and H.~Li.
\newblock Text-dependent speaker verification: Classifiers, databases and
  rsr2015.
\newblock {\em Speech Communication}, 60:56--77, 2014.

\bibitem{li2017deep}
C.~Li, X.~Ma, B.~Jiang, X.~Li, X.~Zhang, X.~Liu, Y.~Cao, A.~Kannan, and Z.~Zhu.
\newblock Deep speaker: an end-to-end neural speaker embedding system.
\newblock {\em arXiv preprint}, 2017.

\bibitem{liu2014speaker}
Y.~Liu, T.~Fu, Y.~Fan, Y.~Qian, and K.~Yu.
\newblock Speaker verification with deep features.
\newblock In {\em IJCNN}, pages 747--753, 2014.

\bibitem{mclaren2015advances}
M.~McLaren, Y.~Lei, and L.~Ferrer.
\newblock Advances in deep neural network approaches to speaker recognition.
\newblock In {\em Acoustics, Speech and Signal Processing (ICASSP), 2015 IEEE
  International Conference on}, pages 4814--4818, 2015.

\bibitem{motiian2017few}
S.~Motiian, Q.~Jones, S.~Iranmanesh, and G.~Doretto.
\newblock Few-shot adversarial domain adaptation.
\newblock In {\em Advances in Neural Information Processing Systems}, pages
  6670--6680, 2017.

\bibitem{nagrani2017voxceleb}
A.~Nagrani, J.~S. Chung, and A.~Zisserman.
\newblock Voxceleb: a large-scale speaker identification dataset.
\newblock {\em arXiv preprint}, 2017.

\bibitem{nakasone2001forensic}
H.~Nakasone and S.~D. Beck.
\newblock Forensic automatic speaker recognition.
\newblock In {\em A Speaker Odyssey-The Speaker Recognition Workshop}, 2001.

\bibitem{park2017using}
S.~J. Park, G.~Yeung, J.~Kreiman, P.~A. Keating, and A.~Alwan.
\newblock Using voice quality features to improve short-utterance,
  text-independent speaker verification systems.
\newblock {\em Proc. Interspeech}, pages 1522--1526, 2017.

\bibitem{sainath2015deep}
T.~N. Sainath, B.~Kingsbury, G.~Saon, H.~Soltau, A.-r. Mohamed, G.~Dahl, and
  B.~Ramabhadran.
\newblock Deep convolutional neural networks for large-scale speech tasks.
\newblock {\em Neural Networks}, 64:39--48, 2015.

\bibitem{snyder2016deep}
D.~Snyder, P.~Ghahremani, D.~Povey, D.~Garcia-Romero, Y.~Carmiel, and
  S.~Khudanpur.
\newblock Deep neural network-based speaker embeddings for end-to-end speaker
  verification.
\newblock In {\em IEEE Spoken Language Technology Workshop (SLT)}, pages
  165--170, 2016.

\bibitem{Soleymani2018multi}
S.~Soleymani, A.~Dabouei, H.~Kazemi, J.~Dawson, and N.~M. Nasrabadi.
\newblock Multi-level feature abstraction from convolutional neural networks
  for multimodal biometric identification.
\newblock In {\em 24th International Conference on Pattern Recognition (ICPR)},
  2018.

\bibitem{Soleymani2018generalized}
S.~Soleymani, A.~Torfi, J.~Dawson, and N.~M. Nasrabadi.
\newblock Generalized bilinear deep convolutional neural networks for
  multimodal biometric identification.
\newblock In {\em IEEE International Conference on Image Processing (ICIP)},
  2018.

\bibitem{variani2014deep}
E.~Variani, X.~Lei, E.~McDermott, I.~L. Moreno, and J.~Gonzalez-Dominguez.
\newblock Deep neural networks for small footprint text-dependent speaker
  verification.
\newblock In {\em IEEE International Conference on Acoustics, Speech and Signal
  Processing (ICASSP)}, pages 4052--4056, 2014.

\bibitem{zhang2017end}
C.~Zhang and K.~Koishida.
\newblock End-to-end text-independent speaker verification with triplet loss on
  short utterances.
\newblock In {\em Proc. of Interspeech}, 2017.

\end{thebibliography}
}
\end{document}